\documentclass[letterpaper,11pt]{article}
\usepackage{graphicx}
\usepackage{float}
\usepackage{rotating}
\usepackage{multirow}
\usepackage{amsfonts,amssymb,latexsym,amsmath,pifont,mathdots,mathtools,natbib}
\usepackage{color}
\usepackage{setspace}
\usepackage{subfig}
\usepackage{url}
\usepackage{epstopdf,epsfig}
\usepackage{tikz}

\newcommand{\kpxx}{\sqrt{K_{xx}}^+}
\newcommand{\kpyy}{\sqrt{K_{yy}}^+}
\newcommand{\kpzz}{\sqrt{K_{zz}}^+}
\newcommand{\Ret}{Re_\tau}

\DeclareRobustCommand{\bcircle}{\protect\tikz{\filldraw circle (2pt)}}
\DeclareRobustCommand{\wcircle}{\protect\tikz{\draw circle (2pt)}}
\DeclareRobustCommand{\bsquare}{\protect\tikz{\filldraw (1pt,1pt) rectangle ++(3pt,3pt)}}
\DeclareRobustCommand{\wsquare}{\protect\tikz{\draw (1pt,1pt) rectangle ++(3pt,3pt)}}

\title{Turbulent boundary layers over streamwise-preferential porous materials}

\author{Christoph Efstathiou and Mitul Luhar\thanks{Email address for correspondence: luhar@usc.edu} \\ 
Department of Aerospace and Mechanical Engineering \\ University of Southern California, Los Angeles, CA 90089}

\begin{document}

\maketitle

\begin{abstract}
Recent numerical simulations indicate that streamwise-preferential anisotropic porous materials have the potential to reduce skin friction in turbulent flows through a similar mechanism to riblets.  This paper reports particle image velocimetry (PIV) measurements made in turbulent boundary layers at $\Ret \approx 360$ over 3D-printed porous substrates exhibiting such streamwise-preferential permeability.  The porous material has normalized streamwise permeability $\kpxx \approx 3.0$  and wall-normal and spanwise permeabilities $\kpyy = \kpzz \approx 1.1$.  This material is flush-mounted into a cutout in the downstream half of a flat-plate boundary layer setup in a water channel facility.  Measurements made at several locations along the porous substrate provide insight into boundary layer development. For fully-developed conditions, the mean profiles show the presence of a logarithmic region over the porous material with similar constants to those found over a smooth wall. A technique that estimates the mean profile at single-pixel resolution from the particle images suggests the presence of an interfacial slip velocity of $U_s^+ \approx \kpxx$ over the porous substrate. Friction velocity estimates obtained from outer layer fits to the mean profile suggest a marginal increase in drag over the porous substrate. PIV measurements show a decrease in the intensity of streamwise velocity fluctuations in the near-wall region and an increase in the intensity of wall-normal velocity fluctuations. These observations are consistent with simulation results, which suggest that materials with $\kpyy > 0.4$ are susceptible to the emergence of spanwise rollers similar to Kelvin-Helmholtz vortices that degrade drag reduction performance. Velocity spectra indicate that such structures emerge in the experiments as well.
\end{abstract}

\section{Introduction}\label{sec:intro}
Functional surfaces have shown significant promise as methods of passive drag reduction in wall-bounded turbulent flows.  Streamwise-aligned \textit{riblet} surfaces have demonstrated drag reductions of up to 10$\%$ in laboratory experiments \citep{walsh1984optimization,bechert1997experiments,bechert2000experiments,garcia2011drag,garcia2011hydrodynamic}. 
More recent work shows that streamwise-preferential porous materials also have the potential to reduce turbulent friction drag through a similar mechanism to riblets \citep{nabil_garcia_dragreduction}.  In particular, numerical simulations suggest that drag reductions of up to $25\%$ may be possible over such anisotropic porous substrates \citep{rosti2018turbulent,gomez2019turbulent}. The work presented in this paper is a step towards testing these predictions in laboratory experiments. 

\subsection{Drag reduction mechanism}\label{sec:mechanism}
The mechanism through which riblets reduce drag can be distilled down to their effect on the near-wall turbulence. By providing greater resistance to turbulent cross-flows compared to the streamwise mean flow, riblets displace the quasi-streamwise vortices associated with the energetic near-wall cycle away from the wall \citep{robinson1991coherent,jimenez_pinelli_1999}, which weakens the vortices and inhibits turbulent mixing near the riblet tips \citep{luchini1991resistance,choi1993direct,jimenez2001turbulent}.  This mechanism can also be understood in terms of the slip lengths perceived by the streamwise mean flow ($l_U^+$) and the turbulent cross-flows ($l_t^+$) below the riblet tips \citep{luchini1991resistance}.  Following standard notation, a superscript $+$ denotes normalization with respect to viscosity $\nu$ and friction velocity $u_\tau$. Since the streamwise mean flow is impeded to a lesser degree than the turbulence cross-flow, it penetrates to a larger distance into the riblet grooves, i.e., $l_U^+ \ge l_t^+$. This offset between the mean flow and the turbulence reduces momentum transfer towards the wall and leads to drag reduction. The length scales $l_U^+$ and $l_t^+$ depend on the shape and size of the riblets, and can be estimated by solving the viscous Stokes flow equations in the streamwise and spanwise directions over the riblets. For small riblets, the drag reduction is expected to be proportional to the difference between the streamwise and transverse slip lengths, $DR = (C_f - C_{f0})/C_{f0} \propto l_U^+ - l_t^+$.  Here, $C_f$ and $C_{f0}$ are the skin friction coefficients over the riblet surface and smooth wall, respectively.  Assuming outer-layer similarity holds \citep[e.g.,][]{flack2007examination}, further away from the wall the effect of small riblets is limited to an upward (drag decrease) or downward (drag increase) shift in the mean velocity profile, which can be quantified in terms of the so-called Hama roughness function $\Delta U^+$ \citep[see e.g.,][]{hama1954boundary,jimenez2004turbulent}.  Previous experiments and simulations show that $\Delta U^+ = m(l_U^+ - l_t^+)$, where $m$ is an $O(1)$ constant \citep[e.g.,][]{bechert1997experiments,garcia2011hydrodynamic,garcia2019control}. Note that the friction drag reduction is related to this shift in the mean profile. For small changes in friction it can be shown that $DR \approx \sqrt{2 C_{f0}} \Delta U^+$.  

Although drag reduction increases initially with increasing riblet size, there is a shape-dependent optimal size beyond which performance degrades. Early studies attributed this deterioration of performance with increasing riblet size to the near-wall turbulence being able to penetrate into the riblet grooves \citep{choi1993direct,lee2001flow}. However, the high-fidelity simulations pursued by \citet{garcia2011hydrodynamic} show that the emergence of energetic spanwise-coherent rollers from a Kelvin-Helmholtz type instability also plays an important role in driving this deterioration of performance.  Recent resolvent-based model predictions also show the emergence of energetic spanwise rollers over riblet surfaces \citep{chavarin2020resolvent}. A compilation of previous results from experiments and simulations suggests that the cross-sectional area of the riblet grooves ($A_g$) is a useful predictive measure for the optimal riblet size beyond which performance degrades \citep{garcia2011drag}. The optimal size across many different riblet shapes is found to be approximately $l_g^+ = \sqrt{A_g}^+ \approx 10.7$ \citep{garcia2011drag,garcia2011hydrodynamic,chavarin2020resolvent}.
Note that spanwise-coherent rollers also appear in numerical simulations and experiments over porous materials \citep{breugem2006influence,rosti2015direct, chandesris2013direct,kuwata2017direct,suga2018anisotropic}. 

Relevant to the present effort, \citet{nabil_garcia_dragreduction} expanded the slip length model proposed by \citet{luchini1991resistance} to flows over anisotropic porous materials characterized by streamwise permeability $K_{xx}$, wall-normal permeability $K_{yy}$, and spanwise permeability $K_{zz}$. This effort shows that $l_U^+ \propto \kpxx$ and $l_t^+ \propto \kpzz$, which implies that $\Delta U^+ \propto \kpxx - \kpzz$.  In other words, turbulent drag reduction may be possible over streamwise-preferential porous materials with $\kpxx > \kpzz$. In addition, linear stability analyses suggest that the maximum achievable drag reduction is limited by the emergence of spanwise rollers as the wall-normal permeability increases.  Results from recent direct numerical simulations (DNSs) support these theoretical predictions.  Specifically, the results obtained by \citet{rosti2018turbulent} and \citet{gomez2019turbulent} show that drag reduction is possible in turbulent flows over anisotropic permeable materials at $\Ret \approx 180$.  In particular, the extensive parametric sweep pursued by \citet{gomez2019turbulent} confirms that the initial drag reduction depends on the difference between the streamwise and spanwise permeabilities, $DR \propto \kpxx - \kpzz$.  In addition, these simulations show drag reductions of up to $25\%$ before the emergence of energetic spanwise rollers leads to a deterioration of performance.  The onset of these Kelvin-Helmholtz type rollers is triggered by wall-normal permeabilities above $\kpyy>0.4$.  

The simulation results described above suggest that streamwise preferential materials with $\kpxx \ge (\kpyy,\kpzz)$ have the potential to reduce drag, as long as the absolute value of the wall-normal permeability remains small, $\kpyy < 0.4$.  However, these simulations employ idealized models for the flow through the porous material (e.g., the Darcy-Brinkman equation) that do not explicitly resolve flow at the pore-scale.  Instead, the effect of the permeable medium is included through the use of simplified effective models involving \textit{bulk} properties such as permeability. Moreover, these prior simulations neglect inertial effects inside the permeable medium that are typically included via the nonlinear Forchheimer term \citep{whitaker1996forchheimer,breugem2006influence}, and they do not account for the variation in bulk properties near the porous interface \citep{lacis_Bagheri_2017}.  Given these simplifying assumptions, it remains to be seen if the trends observed in the numerical simulations hold for physically-realizable materials that have similar bulk permeability. For completeness, we note that previous simulation efforts that employ slip-length or admittance boundary conditions to account for the presence of permeable walls also show the possibility of drag reduction over anisotropic surfaces \citep{jimenez2001turbulent,hahn2002direct,busse_Sandham_2012}.

\subsection{Previous experiments over porous materials}
Experiments involving turbulent flows over porous substrates have thus far been limited by the materials available.  
Experiments motivated by aircraft wings and airfoils have investigated flat plates and bluff bodies with arrays of holes \citep{kong1982turbulent, ruff1972turbulent}. These materials have limited substrate thickness and porosity, and are only permeable in the wall-normal direction.  However, they can have a significant benefit in terms of lift enhancement or stall delay \citep{hanna2019aerodynamic}.
Motivated by environmental flows, several experiments have considered a porous substrate consisting of packed spheres \citep[e.g.,][]{zagni1976channel,blois2020novel}. Packed sphere beds have limited porosity $\epsilon < 0.7$ and are approximately isotropic.  However, they are readily available in different sizes and also allow for refractive index matching. Recent experiments involving index-matched PIV provide useful insight into the relative effects of porosity and roughness on turbulent flows \citep{kim2019piv, kim2020experimental}.  These index-matched experiments also confirm the existence of the so-called amplitude modulation phenomenon observed for smooth wall flows \citep{marusic2010predictive} over, and within, the porous substrates. 

Other experiments studying turbulent boundary layer and channel flows adjacent to porous substrates have utilized high-porosity quasi-isotropic foams \citep{manes2011turbulent, efstathiou2018mean} or meshes with high wall-normal permeability and low streamwise permeability \citep{suga2018anisotropic}. In almost all of these experiments, friction increases substantially over the porous substrates and the velocity fields or energy spectra show the emergence of energetic spanwise-coherent Kelvin-Helmholtz rollers. Laser doppler velocimeter (LDV) measurements made by \citet{efstathiou2018mean} also show evidence of amplitude modulation over high-porosity reticulated foams with permeability Reynolds numbers $Re_k = \sqrt{K} u_\tau/ \nu = \sqrt{K}^+ = 1-9$, where $K$ is a representative scalar permeability.  Measurements made by \citet{manes2011turbulent} using a 2D LDV over high-porosity materials with low surface roughness suggest the existence of a modified logarithmic region in the mean profile with lower von K\'arm\'an constants ranging from $\kappa \approx 0.2$ to $\kappa \approx 0.4$.  A large number of numerical and experimental datasets support the universality of $\kappa = 0.39 \pm 0.02$ \citep{marusic2013logarithmic} for boundary layer flows over impermeable walls.  However, the effects of porous walls on this universality have not been studied in detail. This work does not investigate the divergence of $\kappa$, primarily because $\kappa \approx 0.39$ appears appropriate for the porous materials tested. Very recently, \citet{suga2018anisotropic} used PIV to make measurements in channel flows with bulk Reynolds numbers $Re_b = 900-13600$ over anisotropic materials with $\kpyy > \kpxx$ made of layered and offset meshes. Their PIV measurements in the streamwise-spanwise ($x-z$) and streamwise-wall-normal planes ($x-y$) confirm the emergence of spanwise coherent structures and show drag increases of up to $97\%$. 

None of the experimental efforts described above test the effects of streamwise-preferential permeable materials on turbulent flows. Interestingly, the experiments pursued by \citet{itoh2006turbulent} show drag reductions of up to $12\%$ over porous seal fur, which could be considered a streamwise-preferential porous material.  These experiments tested the effects of both riblets and seal fur in turbulent channel flows at $\Ret \approx 120-600$. Pressure drop measurements showed that the seal fur led to drag reductions that were nearly twice as large as the riblets.  However, profiles of the mean velocity and streamwise fluctuations measured by LDV showed no significant departure from smooth wall profiles. Although the seal fur could be considered a streamwise-preferential material, there are few experiments that involve turbulence measurements over permeable materials with carefully-controlled anisotropy.  To our knowledge, the preliminary channel flow experiments detailed in \citet{chavarin2020resolvent} include the first dataset of turbulent flow over porous materials \textit{designed} to have streamwise preferential permeability. These small-scale experiments tested the effect of 3D-printed porous materials with both $\kpxx > \kpyy$ and $\kpxx < \kpyy$ at $Re_\tau \approx 120$. The streamwise-preferential material with $\kpxx > \kpyy$ did not show a significant departure from smooth wall conditions due to the limited anisotropy. As expected, the material with higher wall-normal permeability triggered the onset of Kelvin-Helmholtz type rollers and led to a significant increase in drag.

\subsection{Contribution and outline}
This paper presents results from laboratory water channel experiments that tested the effect of 3D-printed porous materials with streamwise-preferential permeability on turbulent boundary layer flows at $\Ret \approx 360$. The porous materials were designed to have a cubic lattice microstructure with large openings in the wall normal-spanwise ($y-z$) plane to limit the resistance felt by the mean flow in the streamwise ($x$) direction, i.e., to ensure high streamwise permeability $K_{xx}$.  The openings in the remaining two planes were designed to be smaller to yield lower wall-normal and spanwise permeabilities, $K_{yy}$ and $K_{zz}$, respectively.  Due to fabrication constraints, the exact permeability values did not fall in the range that is expected to yield drag reduction per the simulations of \citet{gomez2019turbulent}.  Specifically, the wall-normal permeability of the fabricated materials is larger than the threshold value above which energetic spanwise rollers are expected to emerge, $\kpyy \approx 0.4$.  Indeed, the measurements reported below show a small increase in skin friction over the anisotropic porous material. Nevertheless, the measurements provide useful insight into the effect of anisotropic porous materials on the mean profile and turbulence statistics. 

The remainder of this paper is structured as follows.  The experimental methods are described in \S\ref{sec:expts}.  This includes details on the flow facility, porous substrate design and fabrication, and diagnostic techniques.  The results are presented and discussed in \S\ref{sec:results}. Flow development over the porous materials is considered in \S\ref{sec:development}.  The mean velocity profile and turbulence statistics for fully-developed conditions at a downstream location over the porous substrate are compared against smooth wall conditions in \S\ref{sec:downstream}.  Changes in velocity spectra are considered in \S\ref{sec:pod-spectra}. Brief concluding remarks are presented in \S\ref{sec:conclusions}.

\section{Experimental methods}\label{sec:expts}
The flow facility and flat plate setup used for the boundary layer experiments are described in \S\ref{sec:facility}. This is followed by a description of the PIV system and processing routines in \S\ref{sec:piv} and the procedure used to estimate the mean profile at single-pixel resolution in \S\ref{sec:single-pixel}.  Design and fabrication of the porous materials is discussed in \S\ref{sec:foams}.  The approach used to estimate friction velocities from the mean profile measurements is presented in \S\ref{sec:utau}.

\subsection{Flow facility and flat plate apparatus}\label{sec:facility}

\begin{figure}
\centering
\includegraphics[width=\textwidth]{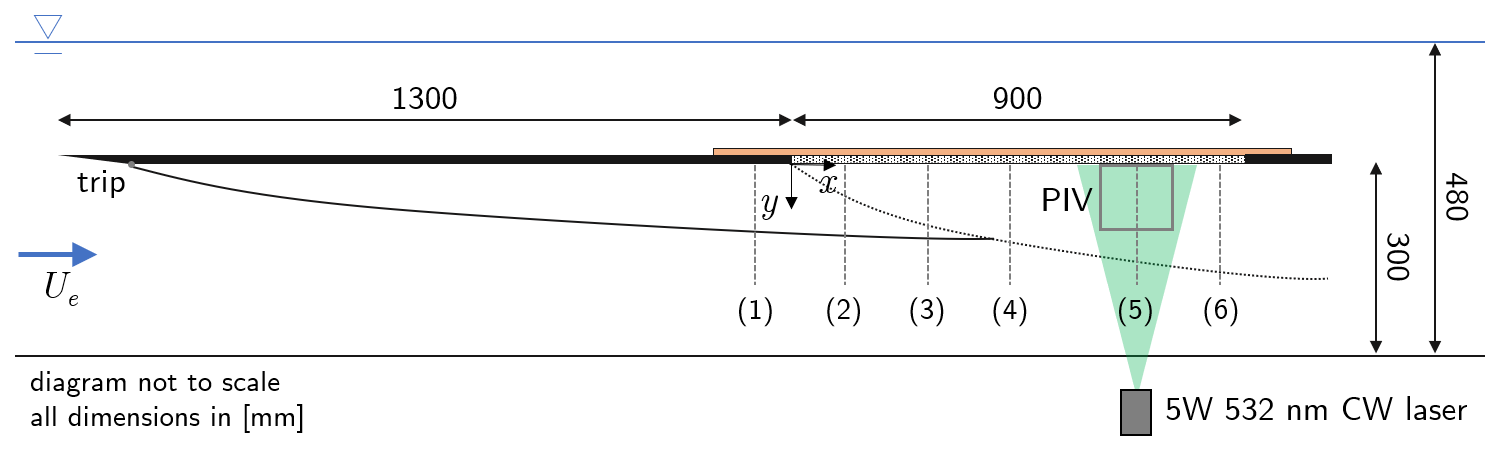}
\caption{Schematic showing experimental setup. The laser, optics and camera system are mounted to a precision traverse (not shown) that can be moved from just upstream of the substrate transition to the end of the plate. The dashed vertical lines indicate the center of the 6 measurement stations at $(x/h)=(-5,3,12,28,44,53)$, where $x=0$ is defined as the start of the porous substrate and $h = 15.4$ mm is substrate thickness.}
\label{fig:diag}
\end{figure}

The experiment utilizes the same flat plate apparatus and water channel facility as described in \citet{efstathiou2018mean}.  However, the experiments are carried out at a lower Reynolds number. A schematic is provided in Fig.~\ref{fig:diag}.  The water channel has a test section of length 762 cm, width 89 cm, and height 61 cm, and is capable of generating free-stream velocities up to 70 cm/s with background turbulence levels $<1\%$ at a water depth of 48 cm. For the present experiments, a 240 cm long flat plate was suspended from precision rails at a height $H=30$ cm above the test section bottom. To avoid free-surface effects, measurements were made below the flat plate. The confinement between the flat plate and bottom of the channel naturally led to a marginal increase ($\leq 4\%$) in the free stream velocity ($U_e$) and slightly favorable pressure gradient along the plate. However, the non-dimensional acceleration parameter, $\Lambda = \frac{\nu}{U_e^2} \frac{d U_e}{dx}$ was of $O(10^{-7})$, suggesting any pressure gradient effects are likely to be mild \citep{PatelPreston1965,de2000reynolds,schultz2007rough}. The water temperature for all experiments was $18\pm 0.5^\circ$C for which the kinematic viscosity is $\nu \approx  10^{-2}$ cm$^2$/s.

A cutout of length 90 cm and width 60 cm, located 130 cm downstream of the leading edge, was used to mount the test surfaces. Smooth and porous surfaces were substituted into this cutout, and mounted flush with the surrounding smooth plate. The porous materials, described in further detail below, were bonded to a solid Garolite\textsuperscript{TM} sheet to provide a rigid structure and prevent bleed flow. Care was taken to minimize gaps and ensure a smooth transition from the solid wall to the porous substrate.  The flow was tripped by a wire of 0.5 mm diameter located 10 cm downstream of the leading edge. In order to generate data at Reynolds numbers similar to the DNS simulations by \citet{gomez2019turbulent}, the channel was run at its lowest practical velocity. The freestream velocity was $U_e = 14.7 \pm 0.1$ cm/s immediately upstream of the porous cutout with $<1\%$ in background turbulence.  The friction Reynolds numbers ranged from $\Ret \approx 290$ to $\Ret \approx 410$ depending on measurement location and substrate type. 

\subsection{2D-2C particle image velocimetry}\label{sec:piv}
Time-resolved velocity fields were acquired using a 2-dimensional, 2-component particle image velocimetry (2D-2C PIV) system in the streamwise-wall normal ($x-y$) plane. The camera, laser, and optical components were mounted to a streamwise traversing cart that moved on precision rails above the water channel.  Measurements were made upstream of the cutout and at five streamwise locations over the porous medium. These measurement locations are termed stations 1-6 (see Table~\ref{table:ut_table}). As a point of comparison, PIV measurements were also made with a smooth-walled insert placed in the cutout.  These baseline measurements we carried out upstream of the cutout (station 1) and at two streamwise locations over the smooth wall (stations 4, 5).  

The flow was seeded with $5$ $\mu$m polyamide seeding particles (PSP, Dantec Dynamics) with specific gravity 1.03.  Illumination was provided by a 5W continuous laser emitting a 532 nm beam, with built-in optics that expanded the beam at $10^\circ$.  The resulting laser sheet was used to illuminate the flow field in the streamwise-wall normal ($x-y$) plane in the middle of the water channel. The laser sheet thickness at the wall location was measured to be less than $0.5\pm0.1$ mm. Images were acquired with a Phantom 410L high-speed camera at a rate of 1000 frames per second.  Images were acquired for 12.5 seconds and subsequently transferred from the camera to the computer. For each flow condition, three runs were acquired approximately 10 minutes apart.  The 99\% boundary layer thickness was $\delta \approx 4$ cm upstream of the cutout, and so the total time series duration of $T = 3\times 12.5 = 37.5$ s translates into approximately $T U_e/\delta \approx 140$ turnover times.  The Phantom 410L camera has a resolution of $1280 \times 800$ pixels with a pixel size of 20 $\mu$m. A 50 mm lens with an aperture of f/1.8 was used to acquire images.  The resulting field of view was approximately 125 mm ($x$) by 170 mm ($y$). The average particle size was roughly $3 \times 3$ pixels. 

The acquired data were processed using standard procedures for 2D-2C time-resolved PIV in DaVis 10 (LaVision GmbH).  The data were processed using a final box size of 16 pixels with 50$\%$ overlap.  The PIV analysis was carried out using image pairs separated by 4 frames to ensure that the particles had a displacement of roughly 4 pixels in the free-stream. Generally accepted vector validation routines were used to identify and remove spurious vectors. At each measurement location, the mean turbulence statistics were averaged in the streamwise direction as well as ensemble averaged over the 3 runs. The wall-normal profiles of mean turbulence statistics were further analyzed and plotted using in-house routines. After processing, the velocity field resolution was $\Delta y = \Delta  x \approx 1.2$ mm.  For a representative friction velocity of $u_\tau \approx 0.7$ cm/s (see Table~\ref{table:ut_table}), this translates into a dimensionless spatial resolution of $\Delta y^+ = \Delta  x^+ \approx 8$. The time step was $\Delta t = 1$ ms, yielding $\Delta t^+ \approx 0.05$.

Standard error estimates for each individual run and correlation uncertainties within DaVis \citep{wieneke2015piv} were small ($\le 0.1\%$ for streamwise velocities, $\le 0.5\%$ for wall-normal velocities).  Correlation values were above 0.9 in the free-stream and 0.7 in the near-wall region.  Averaged over the wall-normal profile, uncertainties for turbulence statistics are estimated to be roughly $0.2\%$ in $U$,  $0.5\%$ for $\overline{u^2}$, $1\%$ for $v^2$, and $3\%$ for $\overline{uv}$. Here, $U$ is the mean velocity in the streamwise direction while $u$ and $v$ are the turbulent velocity fluctuations in the streamwise and wall-normal directions, respectively. An overbar ($\overline{{\cdot}}$) denotes a temporal average for each PIV run, a spatial average in $x$ over the PIV window, and an ensemble average over the 3 runs. Maximum uncertainties in the near-wall region ($y^+ < 30$) are estimated to be $1\%$ in $U$, $2\%$ in $\overline{u^2}$, $2\%$ in $\overline{v^2}$, and $5\%$ in $\overline{uv}$. Variability in $U$ across the three runs for each case was $0.2\%$ in the freestream and $1.5\%$ in the near-wall region. Variability in $\overline{u^2}$ and $\overline{v^2}$ was less than $7\%$ averaged across the profiles.

\subsection{Mean flow estimation at single-pixel resolution}\label{sec:single-pixel}
As noted above, the 2D-2C PIV data were obtained at a spatial resolution of $\Delta y^+ = \Delta  x^+ \approx 8$.  This resolution corresponds to the 8-pixel separation between the 16-pixel interrogation windows with 50\% overlap used for the final pass.  Thus, the actual PIV correlation occurs over windows that are roughly $16\nu/u_\tau$ long in each direction. This leads to substantial spatial averaging in the near-wall region and does not provide sufficient spatial resolution for the evaluation of changes in the near-wall flow.  For example, an estimate of the interfacial slip velocity is impossible at this resolution.

\begin{figure}
\centering
\includegraphics[width=0.8\textwidth]{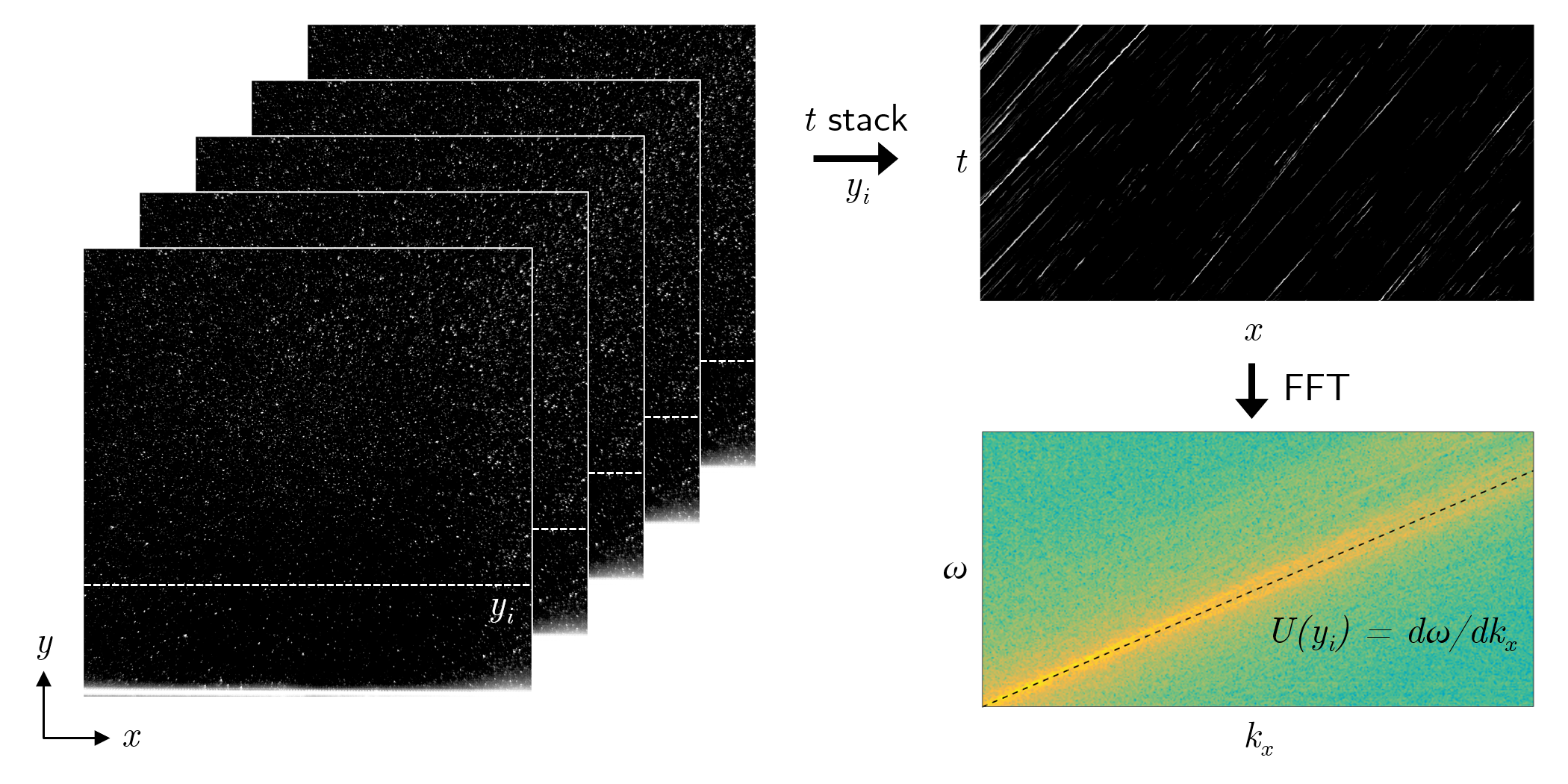}
\caption{Schematic illustration of the single-pixel process. The PIV images are used to create a time-stack showing particle motion in the streamwise direction for each wall normal location, $y_i$. A 2D Fourier transform is used to transform the particle trajectories from $t-x$ space into frequency-wavenumber ($\omega-k_x$) space. A least-squares fit identifies the best estimate for mean particle velocity at each wall-normal location from the group velocity, $U(y_i) = d\omega/d k_x$.}
\label{fig:single}
\end{figure}

To improve spatial resolution, and in particular to evaluate the mean velocity profile near the porous interface, a simple routine was implemented to take advantage of the high temporal resolution of the acquired images, $f_s = 1$ kHz or $f_s^+ \approx 20$. This technique is similar in concept to that proposed by \citet{willert2015high}, who employed single-line correlation on images acquired at 2-7 kHz with a very narrow field of view to resolve the velocity field inside the viscous sub-layer of a turbulent boundary layer at $Re_\tau \approx 240$.  The technique used here is illustrated schematically in Fig.~\ref{fig:single} and described in greater detail below. 

First, the images acquired for PIV are transformed into a time-stack by extracting all the data for a given $y$ location.  In other words, individual rows from each image are stacked together to create a composite image for each wall-normal location ($y_i$) that shows particle motion in the $x-t$ plane. In this composite image, the particle paths appear as diagonal streaks, akin to characteristics. Next, using a 2D fast Fourier transform (FFT), the $x-t$ images are transformed into wavenumber-frequency space ($k_x$ and $\omega$), yielding data similar to that displayed in the bottom right of Fig.~\ref{fig:single}. Finally, the mean velocity $U(y_i)$ for the given wall-normal location is found from a linear least-squares fit to the peak intensity in spectral space, $U(y_i)=d\omega/dk_x$. The procedure outlined above yields an estimate of the mean velocity profile at a wall-normal resolution of $\Delta y \approx 0.15$ mm or $\Delta y^+ \approx 1$.  Note however, that the average particle size is $3\times3$ pixels in the images and so the same particle path can influence mean velocity estimates at 3 different $y$-locations. 

\begin{figure}
\centering
\includegraphics[width=0.5\textwidth]{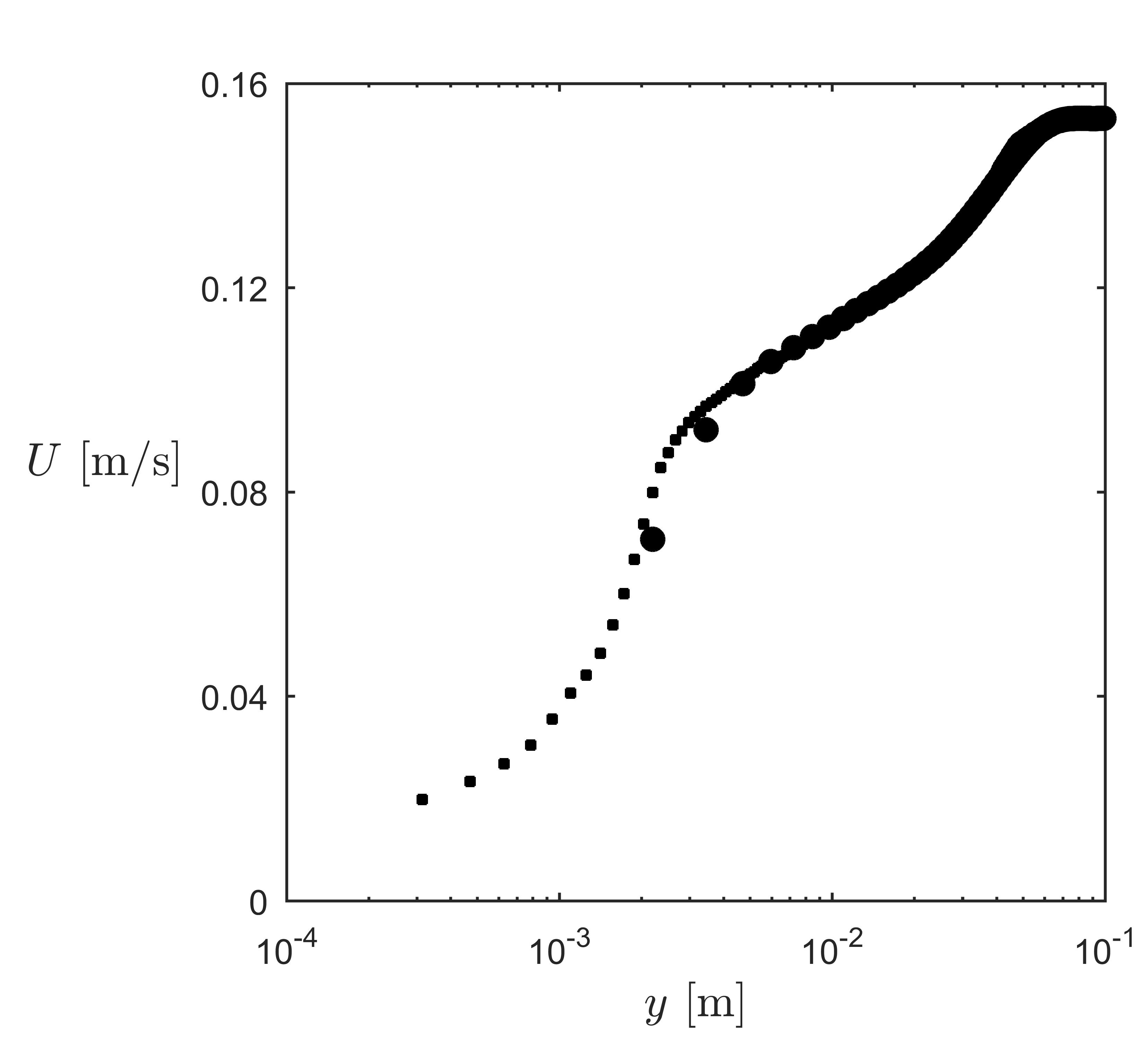}
\caption{Comparison of the mean velocity profile obtained using DaVis (\bcircle) and the single pixel routine (\bsquare) for flow at station 5 over the smooth wall. 
}
\label{fig:single_comp}
\end{figure}

A representative comparison between the higher-resolution profile and the standard PIV profile obtained from DaVis is provided in Fig.~\ref{fig:single_comp}. There is good agreement between the single-pixel profile and the DaVis processed profiles across the boundary layer. Differences between the single-pixel and DaVis profiles are larger in the near-wall region due to the spatial-averaging inherent in the PIV results over the 16-pixel box size in the wall-normal direction. Note that, even though the mean profiles in Fig.~\ref{fig:single_comp} are shown in dimensional terms, it is clear that the single-pixel procedure is able to resolve the mean profile into the viscous sublayer.  Below, we show that the single-pixel mean profiles are also able to provide an estimate for the slip velocity over the porous substrate.  Keep in mind that there is some uncertainty ($\pm 2$ pixels) in estimating the location of the wall in the PIV images for both the smooth wall and porous substrate.



\subsection{Porous substrates}\label{sec:foams}
\begin{figure}
\centering
\includegraphics[width=\textwidth]{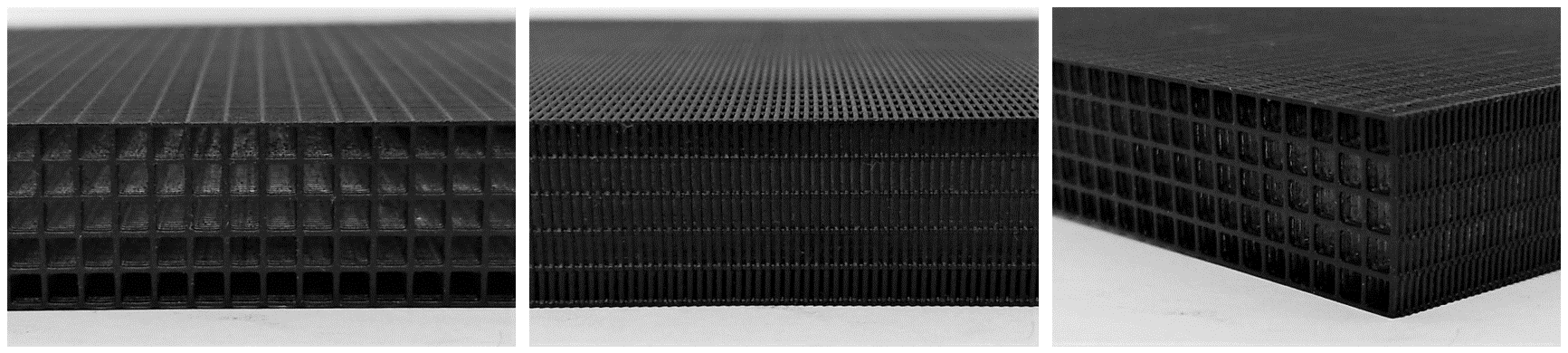}
\caption{
Photographs of the 3D-printed anisotropic porous material.  The streamwise mean flow goes into the page for the image shown on the left. For reference, the total thickness of the porous material is $15.4$ mm.}
\label{fig:foams}
\end{figure}

The anisotropic porous materials were designed and fabricated using the method described in \citet{chavarin2020resolventbased}. Motivated by the simulation results discussed in \S\ref{sec:mechanism}, this procedure was used to generate an anisotropic material that maximized $\kpxx - \kpzz$ while minimizing $\kpyy$ within the fabrication constraints imposed by the 3D printer. The structure of the material consisted of a cubic lattice of rectangular rods with constant cross-section ($d \times d$) and varying spacing in the streamwise ($s_x$), wall-normal ($s_y$) and spanwise ($s_z$) directions. For this experiment, a lattice that maximized pore area in the $y-z$ plane (i.e., normal to the streamwise flow) and minimized the pore areas in the $x-y$ and $x-z$ planes (i.e., facing the spanwise and wall-normal flows) was fabricated using a stereolithographic 3D printer (formlabs Form3). Fabrication constraints (printing resolution, allowable unsupported lengths, resin drainage) limited the maximum anisotropy that could be achieved. 
The minimum pore size was dictated by the printer resolution as the rods fused and the surface became solid if the separation between two rods fell below the laser spot size ($100\mu$m). The maximum pore size was limited by the maximum overhang lengths allowed between rods.  With excessive overhang lengths, the rods sagged and deviated from the design geometry. 
After extensive testing with small samples, a lattice with rod spacings of $s_x = 0.8$ mm and $s_y = s_z = 3.0$ mm and a rod diameter of $d=0.4$ mm was selected for the experiments.  For a representative friction velocity of $u_\tau \approx 7$ mm/s (see Table~\ref{table:ut_table}), the dimensionless rod spacings are $s_x^+ \approx 5.6$ and $s_y^+ = s_z^+ \approx 21$ while the dimensionless rod size is $d^+ \approx 2.8$. This geometry represented a good compromise between generating the desired anisotropic permeability and allowing for reliable manufacturability. Tiles with dimensions of 100mm ($x$) by 15.4 mm ($y$) by 100 mm ($z$) were printed in batches of 5 to reduce manufacturing time. The thickness, $h=15.4$ mm ($h^+ \approx 108$), was selected to allow for 5 full pores in the wall-normal direction. To fill the entire cutout in the flat plate, 90 tiles were manufactured and carefully aligned to preserve streamwise alignment of the pores and minimize gaps. The materials were spray-painted black to reduce reflections from the impinging PIV laser sheet. Sample images of the finished materials are shown in Fig. \ref{fig:foams}. 

The permeability tensor $\mathbf{K}$ of the 3D-printed materials was estimated using Stokes flow simulations run in ANSYS Fluent (Ansys Inc.) following the approach of \citet{zampogna_bottaro_2016}.  To estimate the streamwise permeability, a body force of unit amplitude was imposed in the $x$ direction and the resulting volume-averaged velocity was used to estimate $K_{xx}$ using Darcy's law.  This procedure was repeated with body forces imposed in the $y$ and $z$ directions to estimate $K_{yy}$ and $K_{zz}$, respectively.  The simulations also confirmed zero off-diagonal components in the permeability tensor and so $\mathbf{K} = \mathrm{diag}(K_{xx},K_{yy},K_{zz})$.  The estimated permeabilities for the 3D-printed materials are $K_{xx} =172 \times 10^{-9}$ m$^2$ and $K_{yy} = K_{zz} =22 \times 10^{-9}$ m$^2$.  The porosity of the materials is $\epsilon = 0.87$.  In dimensionless terms, the permeabilities are $\kpxx \approx 3.0$ and $\kpyy = \kpzz \approx 1.1$.  Thus, the difference between the streamwise and spanwise permeabilities yields $\kpxx - \kpzz \approx 1.9$.  Previous theoretical efforts and numerical simulations suggest that the outward shift in the mean profile is expected to be $\Delta U^+ \approx \kpxx - \kpzz$ \citep{nabil_garcia_dragreduction,gomez2019turbulent}.  Thus, the difference between the streamwise and spanwise permeabilities is indicative of a marginal decrease in drag. However, the wall-normal permeability exceeds the threshold identified in \citet{gomez2019turbulent} for the emergence of Kelvin-Helmholtz rollers, $\kpyy \approx 0.4$.

\subsection{Friction velocity estimation}\label{sec:utau}

\begin{figure}
\centering
\includegraphics[width=\textwidth]{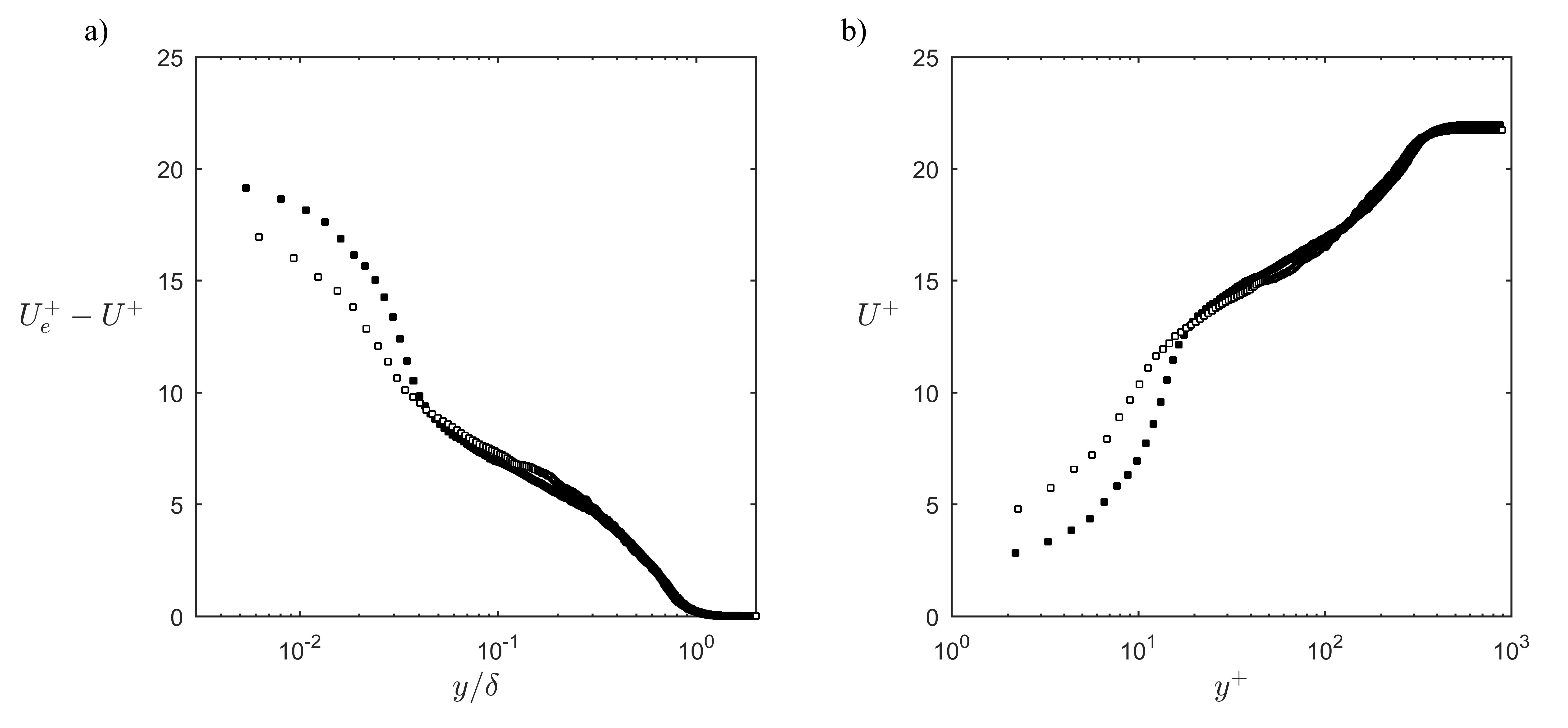}
\caption{The mean velocity profile acquired using the single pixel method is presented in deficit form in panel (a) and in inner-normalized units in panel (b).  Black squares (\bsquare) show measurements made over the smooth wall at station 5 while white squares (\wsquare) show measurements made over the porous substrate at the same location.}
\label{fig:outer}
\end{figure}

Ultimately, this effort seeks to quantify the change in friction over the porous substrate relative to smooth wall values. Unfortunately, the friction drag is not measured directly (e.g., using a force balance). Instead, the friction is estimated indirectly from the friction velocity, $u_\tau= \sqrt{\tau_w/\rho}$, where $\tau_w$ is the shear stress at the wall and $\rho$ is density. The friction coefficient is defined as $C_f =\frac{\tau_w }{1/2 \rho U_e^2} = \frac{2 u_\tau^2}{U_e^2}$. 

For canonical smooth wall, zero pressure gradient turbulent boundary layers, a number of methods have been developed to estimate $u_\tau$. These methods use data from various parts of the boundary layer and fit them to assumed velocity profiles with respective constants \citep[see e.g.,][]{rodriguez2015robust}.  For smooth wall flows, several analytic or implicit formulations exist that can predict the form of the mean profile in the viscous, buffer, and logarithmic regions of the flow \citep[e.g.,][]{clauser1956turbulent,musker1979explicit,kendall2006method}. However, it is unclear if these profiles remain valid over the porous material. Instead, we make use of the logarithmic and wake region data to estimate friction velocity.  In other words, we assume that the outer layer similarity hypothesis holds \citep{townsend1980structure}, such that any changes in the mean profile due to the presence of the porous substrate are restricted to the viscous sublayer and buffer region of the flow.  Outer layer similarity has been validated extensively for rough walls \citep[e.g.,][]{acharya1986turbulent,krogstad1992comparison,flack2007examination}. \citet{monty2016assessment} successfully leveraged outer layer similarity to estimate the friction coefficient for bio-fouled ship hulls. Compared to rough wall flows, \citet{manes2011turbulent} and \citet{efstathiou2018mean} found significant modification to the mean velocity and streamwise turbulence intensity profiles deeper into the boundary layer over high-porosity foams. Nevertheless, both profiles collapsed onto the canonical smooth wall profiles for $y/\delta \gtrsim 0.3$, suggesting that a wake region fit remains applicable here.  Here, $\delta$ is the 99\% boundary layer thickness. This approach also has an additional advantage in that it makes use of logarithmic and wake region data which are more readily available. 


To estimate friction velocity from the logarithmic and wake regions of the flow, we fit the following analytic profile \citep{coles1956law,musker1979explicit,chauhan2009criteria} to the measured mean velocities:
\begin{equation}\label{eq:wake-profile}
U^+ = \frac{1}{\kappa} \log y^+ + B + \frac{\Pi}{\kappa} W(\eta) + \frac{1}{\kappa}\Gamma(\eta). 
\end{equation}
Here, $\kappa$ is the von K\'arm\'an constant, $B$ is the additive constant for the logarithmic region, $\eta = y/\delta$ is the outer-normalized wall normal coordinate, $W(\eta) = 1 - \cos(\pi\eta)$ is the assumed wake function with strength $\Pi$, and $\Gamma = \eta^2 (1-\eta)$. As before, a superscript $+$ denotes normalization with respect to $u_\tau$ and $\nu$.  A least-squares fit to the analytic profile in (\ref{eq:wake-profile}) is used to estimate $u_\tau$, $B$, and $\Pi$ from mean velocity measurements made in the logarithmic region and beyond, i.e., for $y^+ > 30$. The von K\'arm\'an constant is assumed to be constant, $\kappa = 0.39$, for the fitting procedure. Note that we made use of the single-pixel mean profiles for the fitting since the additional data points led to more robust fits.  Figure~\ref{fig:dev} shows the evolution of the fitted parameters $u_\tau$ and $B$ over the porous substrate and smooth wall.  Friction velocity estimates are listed in Table~\ref{table:ut_table}.  To provide uncertainty estimates for the fitted friction velocities, we also attempted fits to just the logarithmic region of the flow as well as the composite profiles proposed by Musker and Spalding \citep{clauser1956turbulent,musker1979explicit,kendall2006method,rodriguez2015robust}. The uncertainty estimate listed in Table~\ref{table:ut_table} is the standard error across these different fits. We also recognize that alternative forms have been proposed for the wake function \citep{chauhan2009criteria}.  A limited sensitivity analysis indicated that the friction velocity estimates obtained were robust to the choice of $W(\eta)$. 

Figure~\ref{fig:outer} shows the mean profiles measured at station 5 over the smooth and porous substrates in outer deficit form (a) and with inner normalization (b). These normalized profiles make use of the fitted friction velocity. The profiles shown in Fig.~\ref{fig:outer}(a) confirm that outer layer similarity holds over the porous materials tested here, at least in the mean velocity profile. The data from the porous case collapse neatly onto the smooth wall data set for $y/\delta \gtrsim 0.1$.  Figure~\ref{fig:outer}(b) shows that the mean profile over the porous substrate departs from the smooth wall profile in the near-wall region. Specifically, the normalized mean velocities are higher over the porous substrate for $y^+ < 10$, which is indicative of a slip velocity at the porous interface. However, both profiles collapse together for $y^+ \gtrsim 30$, which supports the existence of outer layer similarity. Changes in the mean profile over the porous substrate, including the presence of a potential slip velocity, are discussed in greater detail in \S\ref{sec:results}. 

\section{Results and discussion}\label{sec:results}
This section is structured as follows. Boundary layer development over the porous substrate in discussed in \S\ref{sec:development}. Changes in the mean profile and turbulence statistics for fully-developed conditions are considered in \S\ref{sec:downstream}. The effect of the porous substrate on velocity spectra is discussed in \S\ref{sec:pod-spectra}.

\subsection{Boundary layer development}\label{sec:development}

\begin{figure}
\centering
\includegraphics[width=\textwidth]{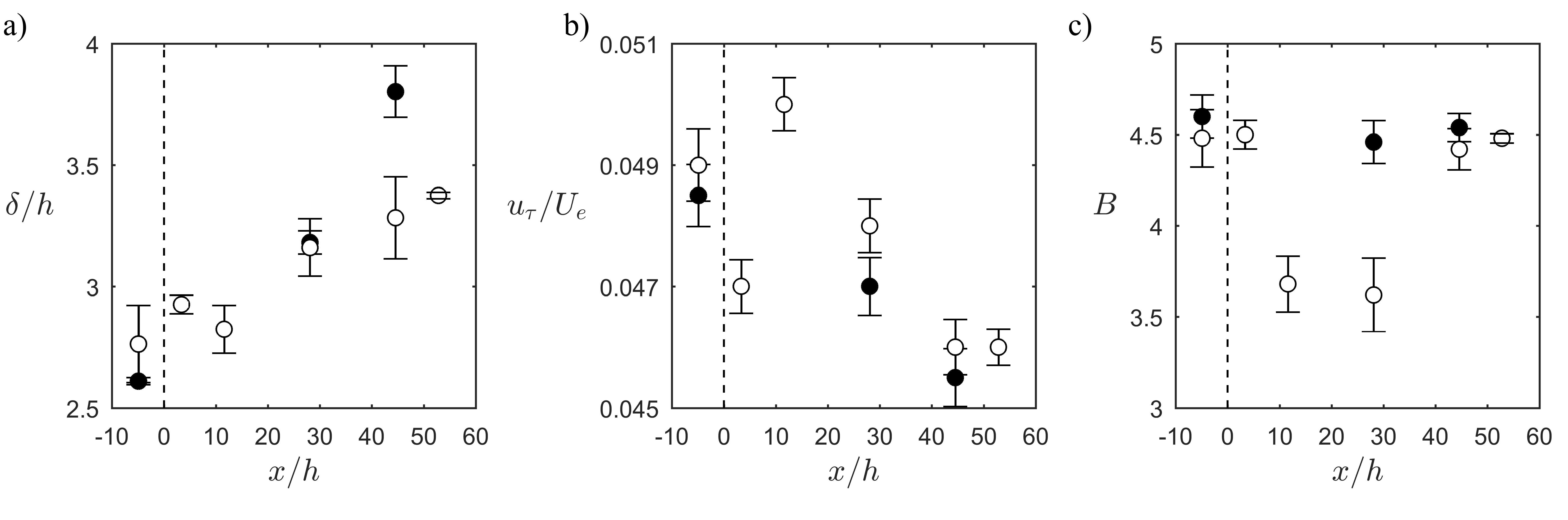}
\caption{Flow development over the porous substrate (\wcircle) and smooth wall (\bcircle).  The evolution of the following parameters is plotted as a function of streamwise location: (a) 99\% boundary layer thickness normalized by substrate height, $\delta/h$; (b) friction velocity normalized by freestream velocity, $u_\tau/U_e$; (c) the additive constant $B$ (\ref{eq:wake-profile}) for the logarithmic region.  The dashed vertical line indicates the transition from the smooth wall to the cutout for the porous substrate.}
\label{fig:dev}
\end{figure}

In this section, we compare flow development over the porous substrate to that over the smooth insert.  For this, we primarily make use of the fitted parameters $u_\tau$ and $B$, the 99\% boundary layer thickness $\delta$, and derived quantities such as the friction coefficient $C_f$.  Note that all of these parameters are obtained from the single-pixel mean profiles.

The first of our measurement locations (station 1) is just upstream of the substrate transition. Subsequent measurement locations (stations 2-6) are located over the cutout into which the porous or smooth inserts are flush-mounted.  PIV measurements were made at all 6 stations for the porous material and at stations 1, 4, and 5 for the smooth wall case.  Figure~\ref{fig:dev} provides insight into flow development over the porous substrate relative to smooth wall conditions.  As expected, the boundary layer thicknesses for both cases agree within uncertainty upstream of the transition.  However, Fig.~\ref{fig:dev}(a) shows that, after an initial perturbation immediately downstream of the transition, the boundary layer thickness grows less rapidly over the porous substrate. For example, at station 5 ($x/h = 44$), the normalized boundary layer thickness over the porous medium is $\delta/h \approx 3.3$ while that over the smooth wall is $\delta/h \approx 3.8$.  This reduction in boundary layer thickness could potentially be attributed to greater flow penetration into the porous substrate. 

Figure~\ref{fig:dev}(b) shows the streamwise evolution of the friction velocity normalized by the freestream velocity, $u_\tau/U_e$.  Again, the friction velocities upstream of the transition agree within uncertainty.  For the smooth wall case, the normalized friction velocities decrease monotonically in the streamwise direction.  However, friction velocities over the porous substrate show some oscillatory behavior at stations 2-3 downstream of the transition.  We attribute this to development effects as the boundary layer adjusts to the new surface condition.  After this initial variability, the normalized friction velocities decrease monotonically over the porous substrate for stations 4-6 ($x/h \ge 28, 44, 53$).  The streamwise evolution of the additive constant for the logarithmic region $B$ shown in Fig.~\ref{fig:dev}(c) is consistent with the friction velocity trends.  Once again, the $B$ estimates agree within uncertainty upstream of the cutout.  Over the smooth wall, the fitted values remain consistent at $B \approx 4.5$.  These smooth wall estimates are a little higher than the typically quoted value of $B \approx 4.3$ for turbulent boundary layer flows \citep{marusic2013logarithmic}, but well within the variability reported in previous literature.  Over the porous substrate, there is a sharp decrease in $B$ at stations 3 and 4, with values around $B \approx 3.6$.  Note that this sharp decrease in $B$ coincides with an increase in $u_\tau$.  For stations 5 and 6 over the porous substrate ($x/h = 44, 53$), the estimated values return to $B \approx 4.5$.  

\begin{table}
	\centering
    \begin{tabular}{cccccc}
        \hline
        Station & $x/h$ & $u_\tau$ [mm/s] & $B$ & $\Ret$ & $C_f$ [$\times 10^3$] \\
        \hline
        1 & -5 &  7.15  &  4.5  &  300  &  4.80 \\
          &    & (7.16) & (4.6) & (290) & (4.70) \\
        2 & 3  &  7.08  &  4.5  &  320  &  4.42  \\
        3 & 12 &  7.58  &  3.7  &  330  &  5.00 \\
        4 & 28 &  7.27  &  3.6  &  350  &  4.61 \\
          &    & (7.05) & (4.5) & (350) & (4.42) \\
        5 & 44 &  7.19  &  4.4  &  360  &  4.23 \\
          &    & (6.98) & (4.5) & (410) & (4.14) \\
        6 & 53 &  6.93  &  4.5  &  360  &  4.23 \\        
        \hline
          &    & $\pm 0.07$ & $\pm 0.1$ & $\pm 10$ & $\pm 0.06$ \\ 
        \hline
    \end{tabular}
	\caption{Estimates for friction-related parameters along the porous substrate. smooth wall values available for measurement stations 1, 4, and 5 are shown in parentheses. Typical uncertainties are shown at the bottom.}
	\label{table:ut_table}
\end{table}

Together, the estimates for $u_\tau/U_e$ and $B$ shown in Fig.~\ref{fig:dev} confirm that the conditions upstream of the cutout are identical (within uncertainty) for the smooth wall and porous substrate experiments. The initial flow development over the porous substrate leads to an increase in friction velocities and a decrease in $B$ (n.b., we recognize that the logarithmic law may not remain appropriate for these non-equilibrium conditions).  However, the flow appears to be fully developed by station 5 located at $x/h = 44$.  This observation is in good agreement with earlier experimental results, which suggest that flow development over porous substrates takes place over a streamwise distance of roughly 40$h$ \citep{efstathiou2018mean}.  At station 5, there is a marginal increase in friction velocity and decrease in $B$ over the porous substrate relative to smooth wall conditions. Estimated values for $u_\tau$, $B$, $\Ret$, and $C_f$ for all measurement locations are listed in Table~\ref{table:ut_table}.  Note that the friction coefficient at station 5 is approximately 2\% higher over the porous medium relative to smooth wall conditions. 

\begin{figure}
\centering
\includegraphics[width=\textwidth]{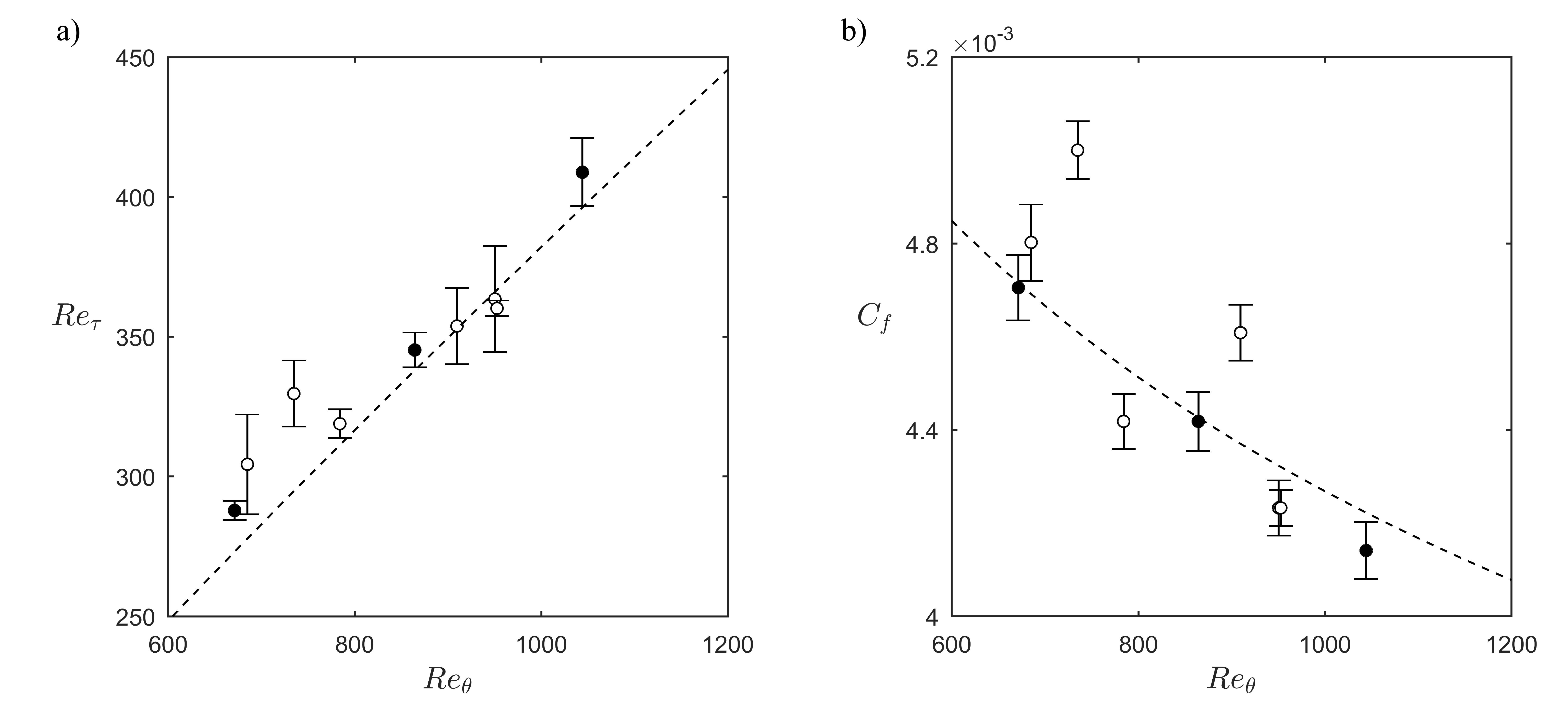}
\caption{Friction Reynolds Number (a) and friction coefficient (b) plotted as a function of the Reynolds number based on momentum thickness,  $Re_\theta$.  Smooth wall values are shown as black circles (\bcircle) while porous substrate values are shown as white circles (\wcircle). Dashed lines show empirical relations from \citet{schlatter2010assessment}: $\Ret = 1.13 Re_\theta^{0.843}$ and $C_f = 0.024 Re_\theta^{-0.25}$. }
\label{fig:cf}
\end{figure}

For completeness, Fig.~\ref{fig:cf} shows friction Reynolds number, $Re_\tau$, and friction coefficient, $C_f$, estimates plotted as as a function of the Reynolds number based on momentum thickness, $Re_\theta$.  Note that the momentum thickness over the porous substrate was estimated only in the unobstructed domain, i.e., this estimate for $\theta$ does not account for flow penetration into the porous medium. Smooth wall estimates for $\Ret$ and $C_f$ agree within uncertainty with previous empirical relations \citep{schlatter2010assessment}.  This provides confidence in the measurement and fitting procedures outlined in the previous sections. 

\begin{figure}
\centering
\includegraphics[width=\textwidth]{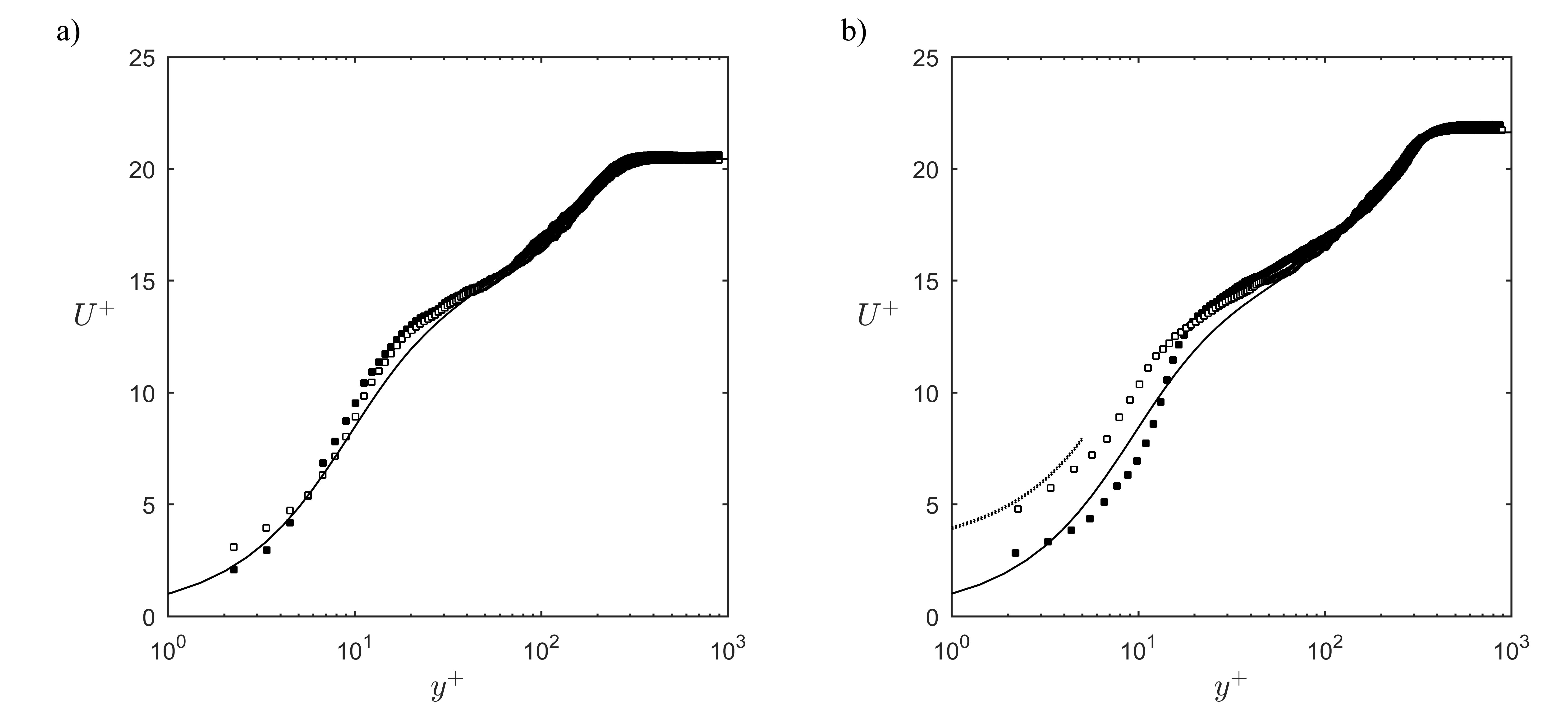}
\caption{Inner-normalized mean velocity profiles measured at station 1 upstream of the cutout (a) and at station 5 ($x/h = 44$) where the flow is fully developed over the porous substrate (b). In both plots, white squares (\wsquare) show measurements from the porous substrate experiments while black squares (\bsquare) show measurements made with the smooth wall insert in place. The solid lines ($-$) show mean profiles obtained in DNS by \citet{schlatter2010assessment} at $\Ret \approx 250$ (a) and at $\Ret \approx 360$ (b).  The dashed line (- -) in panel (b) shows a shifted linear profile of the form $U^+ = \kpxx + y^+$.}
\label{fig:dev-MeanU}
\end{figure}

Finally, Fig.~\ref{fig:dev-MeanU} shows normalized mean velocity profiles collected at station 1 upstream of the cutout and at station 5, where the flow over the porous substrate is expected to be fully developed.  The station 1 mean profiles shown in Fig.~\ref{fig:dev-MeanU}(a) show good agreement between the smooth wall and porous substrate experiments. These profiles show that the single-pixel procedure generates mean profile estimates into the viscous sublayer ($y^+ < 5$).  Moreover, the profiles are in very good agreement with results obtained from DNS at comparable $\Ret$ \citep{schlatter2010assessment}.  The station 5 mean profiles show that the mean velocity over the porous substrate is higher than that over the smooth wall in the buffer region of the flow.  As noted earlier, the mean velocity estimates closest to the porous interface are also indicative of an interfacial slip velocity. These features of the fully-developed mean profile are discussed in \S\ref{sec:downstream} below. In the logarithmic and wake regions of the flow, the smooth wall and porous substrate profiles are in agreement. This observation further supports the existence of a fully-developed condition over the porous substrate for station 5.  

\subsection{Fully-developed flow statistics}\label{sec:downstream}

\begin{figure}
\centering
\includegraphics[width=\textwidth]{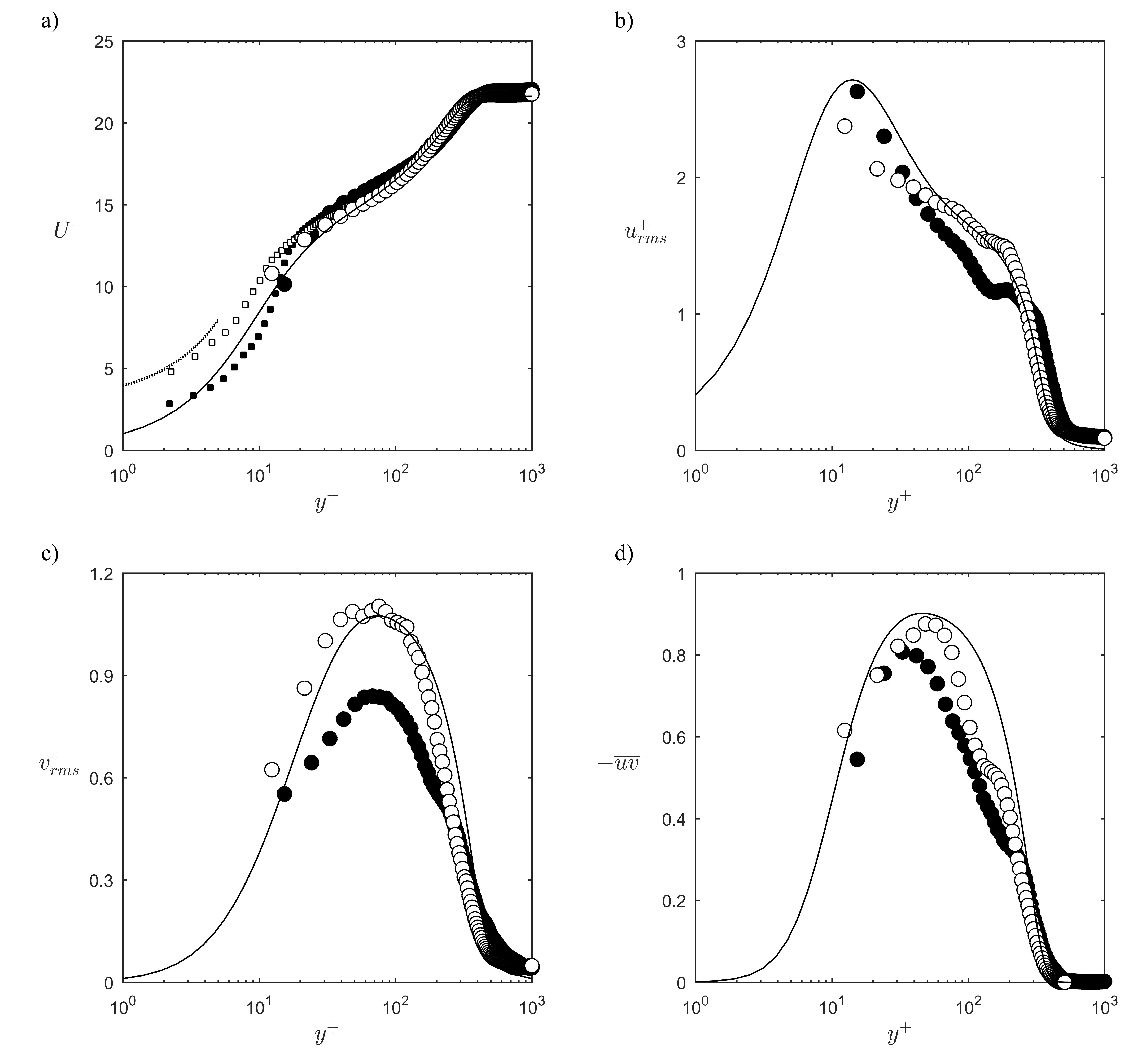}
\caption{Mean turbulence statistics for smooth and porous cases for station 5 at $x/h = 44$. Mean velocity profiles are shown in (a), profiles of the root-mean-square streamwise and wall-normal velocity fluctuations are shown in (b) and (c), respectively.  The Reynolds shear stress profile is shown in (d).  Statistics for the smooth wall and porous substrates are shown as black circles (\bcircle) and white circles (\wcircle) respectively.  The black (\bsquare) and white squares (\wsquare) in (a) show the single-pixel mean profile estimates.}
\label{fig:downstream_stats}
\end{figure}

Figure~\ref{fig:downstream_stats} shows inner-normalized mean statistics obtained from the 2D-2C PIV analysis in DaVis for both the porous and smooth wall cases at station 5 ($x/h = 44$).  As noted in the previous section, the flow over the porous substrate is expected to be fully developed at this location.  Figure~\ref{fig:downstream_stats}(a) shows the mean velocity profiles obtained from the 2D-2C analysis as well as the single-pixel procedure.  In general, the profiles obtained using the two different techniques are in good agreement with one another. For $y^+ > 30$, the mean velocity profiles for both the smooth and porous cases agree well with results from the simulations by \citet{schlatter2010assessment}. For the smooth wall case, the data also compare favorably with the DNS data into the viscous sublayer. There are minor discrepancies between the DNS profile and the single-pixel profile over the smooth wall for $y^+ < 15$.  These discrepancies could be attributed to the $\pm 2$ pixel uncertainty in determining the true location of the wall from the images, which translates into roughly $\pm 2$ viscous units, as well as the uncertainty in the estimate for $u_\tau$.  The mean profile over the porous substrate agrees with the smooth wall profile in the logarithmic and wake regions of the flow.  For $y^+<30$, the mean velocity is higher over the porous medium.  The DaVis profile does not extend into the viscous sublayer of the flow.  However, the mean profile estimated using the single-pixel procedure suggests the presence of a slip velocity with magnitude $U_s^+ \approx \kpxx$.  Specifically, for $y^+ < 5$ the near-wall mean profile over the porous substrate approaches the curve $U^+ = \kpxx + y^+$ (dashed line in Fig.~\ref{fig:downstream_stats}(a)). A slip velocity of $U_s^+ = \kpxx$ is consistent with a slip length of $l_U^+ = \kpxx$ for the mean flow \citep{nabil_garcia_dragreduction}.

Figure~\ref{fig:downstream_stats}(b) shows estimates for the inner normalized root-mean-square (rms) fluctuations in streamwise velocity, $u_{rms}^+ = \sqrt{\overline{u^2}}^+$.  Although the turbulence statistics are not resolved below $y^+ \approx 15$, the streamwise intensity profile over the smooth wall is consistent with the presence of an inner peak at $y^+ \approx 15$.  Further, the streamwise intensity measured at this location is comparable in magnitude to that observed in DNS.  Estimates for the velocity spectra shown in \S\ref{sec:pod-spectra} below confirm that this peak is associated with near-wall structures with frequency $f^+ \approx 0.01$ \citep{robinson1991coherent,jimenez1999autonomous}. Note that the measured streamwise fluctuation intensities over the smooth wall are lower than the DNS values between $y^+ \approx 15$ and $y^+ \approx 200$.  Beyond this location, the DNS and measured profiles show reasonable collapse. In contrast to the mean velocity profiles, the streamwise intensity profile over the porous wall does not collapse onto the smooth wall data until $y^+\approx 200$ or $y/\delta \approx 0.5$. Further, the magnitude of the near-wall peak in streamwise intensity is attenuated by approximately 10\% relative to that for the smooth wall. This observation is consistent with the measurements reported in \citet{efstathiou2018mean} for isotropic foams with comparable wall-normal permeabilities, i.e., $\sqrt{K}^+ = \kpyy \sim O(1)$.  This reduction in peak $u_{rms}^+$ over the porous substrate is also consistent with previous simulation results \citep{breugem2006influence,chandesris2013direct,gomez2019turbulent}.

Profiles for the rms wall-normal velocity fluctuations, $v_{rms}^+ = \sqrt{\overline{v^2}}^+$, are presented in Fig.~\ref{fig:downstream_stats}(c).  Relative to the smooth wall case, the maximum wall-normal fluctuation intensity is roughly 40\% higher over the porous substrate.  Note that the smooth wall profile for $v_{rms}^+$ is also attenuated by roughly 40\% relative to the profile obtained in DNS.  As a result, the profile measured over the porous substrate shows much better agreement with the DNS data.  The quantitative disagreement between the smooth wall measurements and the DNS results can be attributed to the spatial averaging inherent in the PIV analysis algorithm. Recall that the final $16 \times 16$ pixel interrogation windows used in the PIV analyses correspond to boxes that are approximately 16 viscous units in length.  In other words, the PIV measurements cannot properly resolve turbulent flow structures with length scales of $O(10\nu/u_\tau)$.  Since such smaller-scale flow features contribute significantly to the energetic content of the wall-normal velocity fluctuations, the PIV measurements are likely to underestimate $v_{rms}^+$ substantially.  Nevertheless, since both sets of measurements suffer from the same spatial resolution issues, we expect the trends observed in Fig.~\ref{fig:downstream_stats}(c) to remain valid. In other words, the observed increase in $v_{rms}^+$ is likely to hold in measurements made at higher spatial resolution.  Note that the $u_{rms}^+$ profiles do not suffer from the same attenuation because the streamwise velocity fluctuations are typically associated with larger-scale flow features than the wall-normal velocity fluctuations.

The Reynolds shear stress estimates shown in Fig.~\ref{fig:downstream_stats} suffer from the same limitations as wall-normal velocity fluctuations.  It is therefore not surprising that measured profiles of $-\overline{uv}^+$ over both the smooth wall and the porous substrate are lower than the DNS data. The Reynolds stress profile over the porous wall has a peak value that is roughly 10$\%$ higher than over the smooth wall, which is consistent with the increase in $v_{rms}^+$ observed in Fig.~\ref{fig:downstream_stats}(c).

To summarize, the mean statistics shown in Fig.~\ref{fig:downstream_stats} suggest that the porous substrate leads to an interfacial slip velocity of $U_s^+ \approx \kpxx$, a suppression of the near-wall peak in $u_{rms}^+$, and a substantial increase in $v_{rms}^+$ across much of the boundary layer.  These observations are consistent with simulation results obtained by \citet{gomez2019turbulent} over anisotropic porous materials.  Specifically, the conditions tested in the experiments here correspond roughly to cases A5 and A6 in \citet{gomez2019turbulent}.  These cases tested materials with streamwise permeability $\kpxx \approx 2.5 - 3.6$ and wall-normal and spanwise permeabilities $\kpyy = \kpzz \approx 0.7 - 1.0$ in the numerical simulations. Both substrates led to an increase in skin friction relative to smooth wall conditions. \citet{gomez2019turbulent} attributed this increase in skin friction, and the associated changes in turbulence statistics, to the emergence of energetic spanwise rollers with streamwise wavelengths $\lambda_x^+ \approx 100-400$.  We consider the emergence of such rollers by evaluating frequency spectra for the velocity fluctuations in the following section. 


\subsection{Velocity spectra}\label{sec:pod-spectra}

\begin{figure}
\centering
\includegraphics[width=\textwidth]{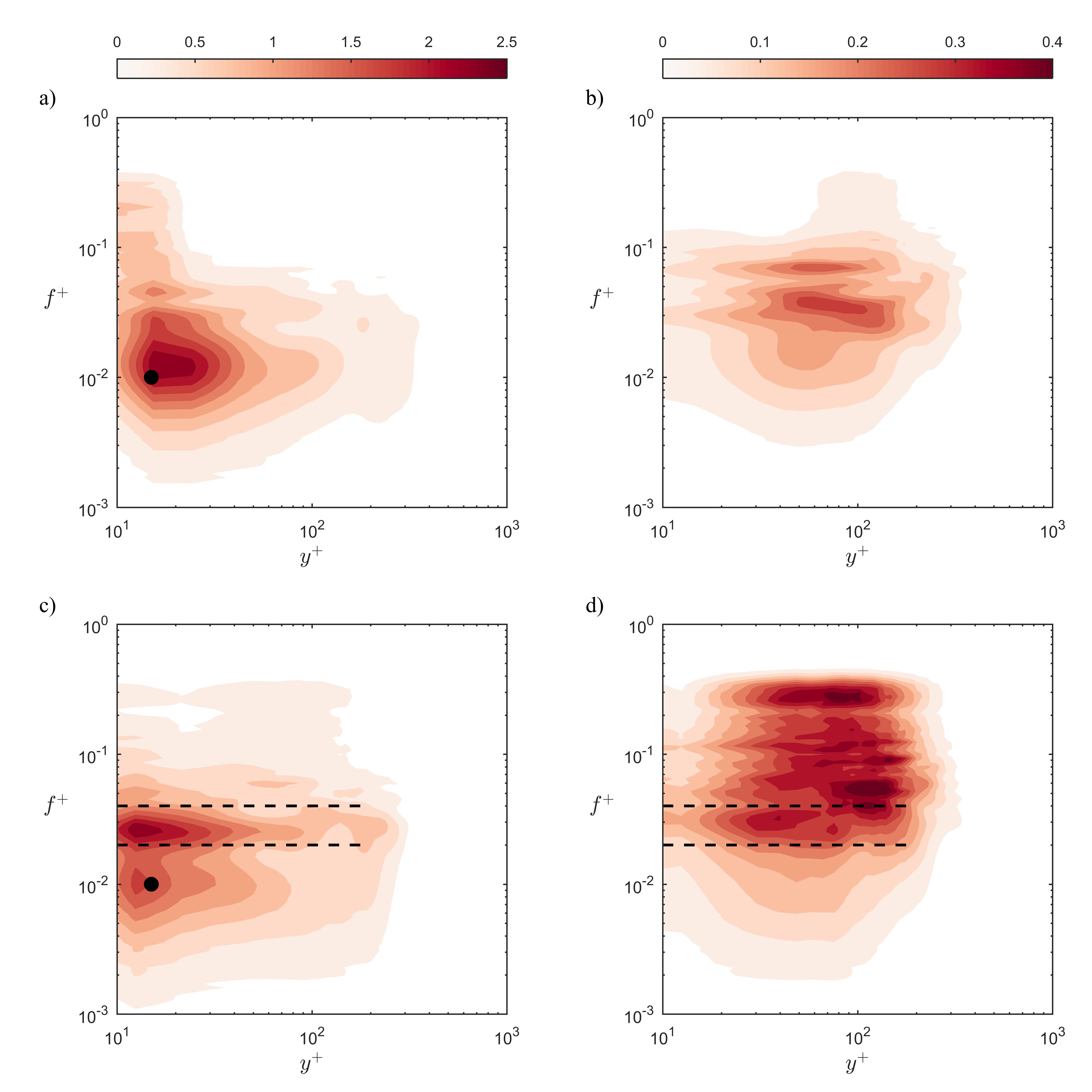}
\caption{Premultiplied spectra at station 5 for the streamwise velocity fluctuations $f^+ E_{uu}^+$ (a,c) and wall-normal velocity fluctuations $f^+ E_{vv}^+$ (b,d).  Smooth wall results are shown in (a,b) and porous substrate results are shown in (c,d). The black circles (\bcircle) in (a,c) label a frequency of $f^+ = 0.01$ at $y^+ = 15$, which corresponds roughly to structures associated with the energetic near-wall cycle.}
\label{fig:spectra}
\end{figure}

Figure~\ref{fig:spectra} shows premultiplied frequency spectra for the streamwise and wall-normal velocity fluctuations, $f^+ E_{uu}^+$ and $f^+ E_{vv}^+$, over the smooth wall and porous substrate at station 5 ($x/h = 44$).  Here, $f$ is frequency while $E_{uu}$ and $E_{vv}$ are the spectral densities for the streamwise and wall-normal velocity fluctuations. These spectra were computed from the DaVis time series of $u$ and $v$ using Welch's algorithm in Matlab (Mathworks, Inc.).  The computed spectra are somewhat noisy --- particularly for the wall-normal velocity fluctuations --- indicating that the acquisition time may not have been long enough for complete convergence.  However, these spectra do provide additional insight into the change in turbulence characteristics over the porous substrate.

The smooth wall velocity spectra shown in Fig.~\ref{fig:spectra}(a) for $u$ and in Fig.~\ref{fig:spectra}(b) for $v$ are broadly consistent with previous observations in wall-bounded turbulent flows \citep{jimenez2008turbulent,jimenez2010turbulent,krishna2020reconstructing}.  Specifically, Fig.~\ref{fig:spectra}(a) shows the presence of a distinct peak in $f^+ E_{uu}^+$ centered near $f^+ \approx 0.01$ and $y^+ \approx 15$.  This peak corresponds to streak-like structures with streamwise wavelength $\lambda_x^+ = U^+/f^+ \approx 10^3$ that are associated with the energetic near-wall cycle \citep{robinson1991coherent,smits2011high}.  Note that $U^+ \approx 10$ at this wall-normal location.  Figure~\ref{fig:spectra}(b) shows that the peak in $f^+ E_{vv}^+$ is centered further away from the wall ($30 < y^+ \lesssim 100$) and at higher frequencies, $f^+ \approx 0.03 - 0.08$.  In other words, the wall-normal velocity spectra are dominated by structures in the logarithmic region of the flow that have streamwise length scales $\lambda_x^+ = U^+/f^+ \sim O(10^2)$, which is in agreement with prior results from numerical simulations \citep{jimenez2001turbulent,krishna2020reconstructing}. 

Figure~\ref{fig:spectra}(c) shows some important changes to $f^+ E_{uu}^+$ over the porous substrate.  Although the usual near-wall peak remains, another region of high energy emerges for structures with $f^+ \approx 0.02 - 0.04$ (see dashed lines).  This region of high $f^+ E_{uu}^+$ extends from the lowest measurement location at $y^+ \approx 10$ out to $y^+ \approx 200$.  We suggest that this is the spectral footprint of the energetic spanwise rollers responsible for drag increases in numerical simulations \citep{gomez2019turbulent}.  Assuming a velocity scale of $U^+ \approx 10$, the frequency range $f^+ \approx 0.02 - 0.04$ translates into structures with streamwise wavelength $\lambda_x^+ \approx 250-400$.  These length scales are in the range identified by \citet{gomez2019turbulent}. The large wall-normal extent is also consistent with previous simulation results, which show that the interfacial Kelvin-Helmholtz type rollers that emerge over porous substrates can extend out into the (nominally) logarithmic region of the flow \citep[e.g.,][]{breugem2006influence}.  Note that a region of high spectral content for $f^+ \approx 0.02 - 0.04$ is also evident in the spectra for wall-normal velocity fluctuations over the porous substrate (see Fig.~\ref{fig:spectra}(d)). However, this region is not as distinct since $f^+ E_{vv}^+$ values are generally elevated over the porous substrate relative to smooth wall conditions. 

\section{Conclusions}\label{sec:conclusions} 
This paper reports some of the first turbulence measurements made in boundary layers over streamwise-preferential porous materials that have demonstrated drag reduction capabilities in recent modeling and simulation efforts \citep{nabil_garcia_dragreduction,rosti2018turbulent,gomez2019turbulent}.  Models developed in these prior studies show that materials with high streamwise permeability and low spanwise permeability (i.e., materials with $\kpxx - \kpzz > 0$) are promising candidates for passive drag reduction.  Driven by these predictions, we designed and 3D-printed a porous substrate with normalized permeabilities $\kpxx \approx 3$ and $\kpyy = \kpzz \approx 1.1$.  

Results presented in \S\ref{sec:development} show that the initial development over the porous substrate takes place over a streamwise distance of $x/h \approx 40$ which is similar to the development length observed in previous experiments over high-porosity foams \citep{efstathiou2018mean}.  For fully developed conditions, indirect friction estimates show that the 3D-printed porous substrate led to a small ($<5\%$) increase in drag.  Despite the drag increase, the experimental measurements are in broad agreement with previous simulation results. For instance, mean profile estimates obtained at single-pixel resolution (see \S\ref{sec:single-pixel}) indicate the presence of a slip velocity over the porous substrate that is consistent with theoretical predictions, $U_s^+ \approx \kpxx$.  Further, PIV-based measurements of turbulence statistics (\S\ref{sec:downstream}) and velocity spectra (\S\ref{sec:pod-spectra}) indicate that the observed drag increase can be attributed to the emergence of energetic spanwise rollers resembling Kelvin-Helmholtz vortices. The simulation results of \citet{gomez2019turbulent} show that such rollers emerge in turbulent flows over anisotropic porous substrates once the wall-normal permeability exceeds $\kpyy \approx 0.4$.  The wall-normal permeability of the material tested here exceeds this threshold value. 

Together, these observations suggest that streamwise-preferential porous materials continue to be promising candidates for passive drag reduction in wall-bounded turbulent flows. At the very least, the results presented in this paper suggest that streamwise-preferential porous substrates could be used for other flow control applications (e.g., to enhance heat transfer) with minimal frictional penalties. 

Of course, the experiments reported here do have some important shortcomings. For example, the PIV results shown in \S\ref{sec:downstream} do not include any turbulence measurements below $y^+ \approx 10$.  A more complete characterization of the interfacial turbulence requires additional measurements made at higher spatial resolution. Such measurements would also help evaluate whether roughness effects due to the presence of the rods of size $d^+ \approx 2.8$ at the porous interface are important. 

Finally, keep in mind that the material tested here has relatively large pore openings (rod spacings of $s_x^+ \approx 5.6$ and $s_y^+ = s_z^+ \approx 21$).  Such large pore openings allowed us to create the desired anisotropy in permeability.  However, this also means that inertial effects are likely to be important for the pore-scale flow. The numerical simulations of \citet{rosti2018turbulent} and \citet{gomez2019turbulent} make use of idealized models for flow within the permeable substrate. These models do not account for interfacial roughness or inertial effects in the porous medium. Geometry-resolving simulations similar to those pursued by \citet{kuwata2017direct} are needed to evaluate whether streamwise-preferential porous materials are capable of drag reduction once inertial effects become important.

\section*{Acknowledgments}
This paper is based on work supported by the Air Force Office of Scientific Research under awards FA9550-17-1-0142 (program manager Dr. Gregg Abate) and FA9550-19-1-7027 (program manager Dr. Douglas Smith). The authors would like to thank Andrew Chavarin for the numerical permeability estimates listed in \S\ref{sec:foams} and Aidan Rinehart for detailed discussions about PIV processing routines.

\bibliographystyle{plainnat}
\bibliography{Efstathiou2020}

\end{document}